\documentclass{aastex}

\received{July 28, 2000}
\accepted{September 18, 2000}

\slugcomment{To be published in the Astrophysical Journal (Letters)}

\shorttitle{Magnetic Fields in Cygnus A}
\shortauthors{Wilson, Young \& Shopbell}

\begin{document}

\title{Chandra Observations of Cygnus A: Magnetic Field Strengths in the Hot
Spots of a Radio Galaxy}

\author{A. S. Wilson\altaffilmark{1}, A. J. Young}
\affil{Astronomy Department, University of Maryland, College Park,
MD 20742; wilson@astro.umd.edu, ayoung@astro.umd.edu}

\and 

\author{P. L. Shopbell}

\affil{Department of Astronomy, Mail Code 105-24,
California Institute of Technology, Pasadena, CA 91125; pls@astro.caltech.edu}


\altaffiltext{1}{Adjunct Astronomer, Space Telescope Science Institute,
3700 San Martin Drive,
Baltimore, MD 21218; awilson@stsci.edu}


\begin{abstract}
We report X-ray observations of the powerful radio galaxy Cygnus A with the
Chandra X-ray Observatory. This letter focuses on the radio hot spots, all
four of which are detected in X-rays with a very similar morphology to their
radio structure. X-ray spectra have been obtained for the two brighter hot
spots (A and D). Both are well described by a power law with photon index
$\Gamma$ = 1.8 $\pm$ 0.2 absorbed by the Galactic column in the direction of
Cygnus A. Thermal X-ray models require too high gas densities and may be ruled
out. The images and spectra strongly support synchrotron self-Compton models
of the X-ray emission, as proposed by Harris, Carilli \& Perley on the basis
of ROSAT imaging observations. Such models indicate that the magnetic field
in each of the brighter hot spots is 1.5 $\times$ 10$^{-4}$ gauss, with an
uncertainty of a few tens of percent. This value is close to the equipartition
field strengths assuming no protons are present. The
possibility that the X-rays are synchrotron radiation is briefly discussed, but 
not favored. We speculate that production of the $\gamma$ $\sim$ 10$^{7}$
electrons necessary for X-ray synchrotron radiation from hot spots is
inhibited when the external gas density is high, as is the case when the
radio galaxy is within a cooling flow.

\end{abstract}




%

\newpage
\section{INTRODUCTION}

The spectral emissivity of optically thin synchrotron radiation is proportional
to n$_{\rm e0}$B$^{\rm (p + 1)/2}$, where n$_{\rm e0}$
is the constant in the relativistic
electron number density spectrum
(n$_{\rm e}(\gamma)d\gamma$ = n$_{\rm e0}{\gamma^{-\rm p}} \rm d\gamma$,
$\gamma$ = E/m$_{\rm e}$c$^{2}$) and B is the
magnetic field strength. Observations of synchrotron sources allow this
product to be determined, but not n$_{\rm e0}$ and 
B separately. It is common practice
to quote the equipartition magnetic field, B$_{\rm eq}$, the field
for which the
energies in relativistic particles and magnetic field are equal; this field is
close to that at which the total energy in relativistic particles plus magnetic
field is a minimum. This inability to obtain a direct measurement of the
magnetic field strength has seriously hampered progress in understanding cosmic
synchrotron sources.

The relativistic electrons will inevitably scatter any photons in the source
through the inverse Compton process. The emissivity of inverse Compton 
radiation depends on the electron energy spectrum and the radiation spectrum,
the latter being an observable quantity for an isotropic source.
Detection of inverse Compton radiation thus permits n$_{\rm e0}$
to be measured and
hence B from the synchrotron spectrum. If the photons being scattered are the
synchrotron radiation itself, the resulting emission is called
synchrotron self-Compton (SSC) radiation.

For a typical extended non-thermal radio source, much of the inverse Compton
emission is expected to be radiated at X-ray energies. Unfortunately, the
existence of other mechanisms that generate X-rays - thermal and synchrotron
emissions - and the poor spatial and spectral resolutions of previous X-ray
observatories has precluded the use of this method to measure magnetic
field strengths in either supernova remnants or extragalactic radio sources.

In this paper, we present the first results of a study of the nearby
(z = 0.0562), powerful radio galaxy Cygnus A with the Chandra X-ray Observatory.
We focus on the X-ray emission of the radio hotspots. Previously,
Harris, Carilli \& Perley (1994, hereafter HCP) detected soft X-ray emission 
from the dominant western and eastern hot spots (termed A and D by
Hargrave \& Ryle 1974) with the High Resolution Imager of the ROSAT
Observatory. They compared the measured X-ray fluxes with predictions of SSC
models and argued that the magnetic fields in the hot spots are quite close to 
the equipartition values, calculated assuming no contribution from relativistic
protons. The angular resolution of the ROSAT HRI was $\simeq$ 
5$^{\prime\prime}$, so the point spread function included both weaker
radio structures (such as the fainter hot spots B and E - cf. Perley, Dreher
\& Cowan 1984) and thermal emission from the cluster gas. More significantly,
the detector had almost no spectral resolution so HCP were unable to measure
the X-ray spectra of the hot spots. The Chandra observations overcome these
limitations. The sub arc second angular resolution and high sensitivity
of Chandra have allowed detection and spatial resolution of X-ray emission from
all four hot spots. We also report the X-ray spectra of the two brightest
hot spots (A and D), finding excellent agreement with the predictions of the
SSC model. This agreement strongly favors the SSC model, but does not rule
out a contribution from X-ray synchrotron emission, as we briefly discuss.
We use H$_{0}$ = 50 km s$^{-1}$ Mpc$^{-1}$ and q$_{\rm 0}$ = 0 throughout.
Other aspects of the Chandra results on Cygnus A will be discussed elsewhere.

\section{OBSERVATIONS AND REDUCTION}

Cygnus A was observed by the Chandra X-ray Observatory on May 21 2000
(sequence number 700032, obsid 360) using the Advanced CCD Imaging
Spectrometer (ACIS) spectroscopic array. The nucleus was centered
20$^{\prime\prime}$ in the --Y direction from the location of best focus on
chip S3. All of the regions of radio emission from Cyg A were imaged on
S3. The total good time interval was 34.7 ksecs taken with the default frame
time of 3.2 secs. 
The data extraction and
analysis have been performed using version 1.1.3 of the CIAO software and
version 11.0 of XSPEC.
A new level 2 events file, with the correct gain map
(acisD2000-01-29gainN0001.fits), was made from the events file supplied by 
the Chandra Science Center. The data were inspected for bad aspect 
and high background times, but none were found. 
The response matrix file
and ancillary response file were initially obtained from calibration data
obtained with the chip at -110C. The spectra were recalculated with
calibration data for a temperature of -120C (the temperature during
the observations) when these became available, but were insignificantly
different from those obtained with the -110C calibration. The counts used 
to obtain the spectra of hot spots A and D were taken from circular regions of
radii 2\farcs3 and 2\farcs2, respectively, with background taken from annular
regions of width $\simeq$ 2$^{\prime\prime}$ concentric with the source region.

\section{RESULTS} 

\subsection {Morphology}

Fig. 1 is a grey scale plot of the X-ray emission from the region of the
radio source. The Chandra astrometry shows that the nuclear X-ray source
agrees with the nuclear radio source to within 1\farcs4. When the X-ray
and radio nuclei are aligned, the compact X-ray sources at the SE and NW
edges of Fig. 1 coincide with the corresponding
radio hot spots to within 0\farcs5. Fig 2
shows X-ray contours on a grey scale of a 6 cm image with
resolution 0\farcs35 (Perley, Dreher \& Cowan 1984; Carilli \& Barthel 1996)
in the vicinities of the western 
(A and B) and eastern (D and E) hot spots. 

As may be seen in Fig. 2, the extents and morphologies of the X-ray and radio
hot spots are very similar, with the directions of elongation agreeing to
within a few degrees. Despite the somewhat lower resolution of the X-ray image,
it is clear that the X-ray emission comes from essentially the same region
as the radio
emission in each hot spot.

\subsection {Spectra}

Hot spots B and E are too weak to obtain reliable X-ray spectra, so we focus
on A and D. The results of modelling each spectrum with an absorbed power
law are shown in Table 1. In both cases, the absorbing column is
N$_{\rm H}$ = 3.3 $\times$ 10$^{21}$ cm$^{-2}$, in excellent agreement
with the Galactic column (3.3 $\times$ 10$^{21}$ cm$^{-2}$, HCP) in the
direction of Cygnus A. The photon indices are similar at
$\Gamma$ = 1.8 $\pm$ 0.2. Alternatively, the spectra may be well described
by a Raymond-Smith thermal plasma model with temperatures 4.9 keV (A) and
6.0 keV (D).

\subsection {X-ray Emission Mechanism}

A thermal model requires a density of 0.5 cm$^{-3}$ in each hot spot. This
value is 10$^{3}$ times larger than the upper limit to the internal density
in the hot spots from the absence of Faraday depolarization (Dreher, Carilli
\& Perley 1987). It is also hard to understand how such a high density could
be produced in the hot spots given that the density of the intracluster medium
near them is only $\simeq$ 0.01 cm$^{-3}$ (HCP; Reynolds \& Fabian 1996).
We conclude that a thermal model for the X-ray emission of the hot spots
is untenable. 

In view of the success of a SSC model in reproducing the intensity
of the soft X-ray emission (HCP), it is natural to check whether the model
can also reproduce the X-ray spectra. From the radio spectra of the hot spots
(Carilli et al. 1991), we first calculated the internal radiant energy
density in the hot spots, modelled as uniformly emitting spheres,
from $\epsilon_{\rm R}$ = 3L$_{\rm R}$R/4cV, where L$_{\rm R}$ is the total
radio luminosity, R and V are the radius and volume of the hot spot, and
c is the speed of light. The results are $\epsilon_{\rm R}$ $\simeq$
3 $\times$ 10$^{-11}$ erg cm$^{-3}$ for each hot spot. These values are
$\simeq$ 100 times larger than the energy density of the microwave
background ($\epsilon_{\rm M}$ $\simeq$ 4 $\times$ 10$^{-13}$ erg cm$^{-3}$),
showing that an SSC model is indeed appropriate.
The radiant energy densities in the hot spots
are, however, $\simeq$ 100 times smaller than that in the magnetic field
($\epsilon_{\rm B}$ $\simeq$ 3 $\times$ 10$^{-9}$ erg cm$^{-3}$) assuming
equipartition and no relativistic
protons. Thus the rate of energy loss by the electrons to synchrotron
radiation will be $\simeq$ 100 times
larger than to SSC radiation.

To calculate the SSC spectra, we have
used the code of Band \& Grindlay (1985, 1986)
which assumes spherical geometry. Following Carilli et al. (1991), the radio
spectrum of each hot spot was modelled as a broken power law with spectral index
$\alpha$ = 0.55 (hot spot A, S $\propto$ $\nu^{-\alpha}$) and 0.50
(hot spot D) below the break frequency and 1.05 (A) and 1.0 (D) above it.
Such a change in slope of 0.5 is expected in models of continuous injection of
electrons accompanied by synchrotron losses (Kardashev 1962). The emitting
region was taken to be a uniform sphere in each case.

The results of the modelling are given in Table 2 and compared with the
Chandra spectra in Figs 3a and b. As may be seen, the predicted SSC radiation
is in excellent agreement with the Chandra-observed spectrum for a magnetic
field of 1.5 $\times$ 10$^{-4}$ gauss in each hot spot, in good agreement
with the results of HCP based on the X-ray intensity. This value may be compared
with the equipartition values of 2.8 $\times$ 10$^{-4}$ gauss (hot spot A)
and 2.5 $\times$ 10$^{-4}$ gauss (D), calculated assuming no relativistic
protons, the broken power-law spectra of Carilli et al. (1991), low
frequency cut-offs at 10 MHz and high frequency cut-offs at 400 GHz. The 
calculated field is insensitive to the precise cut-off frequencies.
The SSC model predicts spectral steepening towards higher energies within the
Chandra band. We have searched for this effect, finding a hint of
a larger value of $\Gamma$ in the 2.5 - 6 keV band than in the 0.7 - 2.5
keV band for hot spot D. However, this difference in photon indices is not 
significant.

Uncertainty in the magnetic field obtained with the SSC model results from
a number of factors, including a) the idealisation of the hot spots as
uniform spheres of known radii, b) the errors in the Chandra-measured
spectra, and c) modelling the radio spectrum of each hot spot as a
broken power law, with sharp changes in the slopes of both the electron
energy
and the synchrotron radiation spectra at the break energy and
frequency. Comparison
of the formulae for synchrotron and inverse Compton radiation indicates that
changing the volume by a factor of 2 changes B by $\sim$ 30\%. The normalisation
of the Chandra spectra are uncertain by $\simeq$ 17 - 28\% (Table 1).
An error of 25\% in SSC flux changes B by $\simeq$ 15\%. Lastly, the idealised
treatment of the electron energy and synchrotron spectra around the
break is estimated to contribute $\lesssim$ 10\% uncertainty. The precise
locations and shapes of the high energy cut-offs have little effect on the SSC
spectra in the Chandra band. We conclude that the error in the magnetic
fields derived from the SSC model amounts to a few tens of percent.

\section{CONCLUDING REMARKS}

The X-ray spectra of the two brightest radio hot spots (A and D) of Cygnus A
are in excellent agreement with an SSC model in which the magnetic field is
1.5 $\times$ 10$^{-4}$ gauss. This value is close to the equipartition
fields of 2.8 $\times$ 10$^{-4}$ and 2.5 $\times$ 10$^{-4}$ gauss for hot spots
A and D respectively, calculated assuming no relativistic protons are present
(i.e. the ratio of the energy density in relativistic protons to that in
electrons, K = 0).
The most straightforward interpretation is that the relativistic gas is an
electron-positron plasma and is close to equipartition with the magnetic
field. If, on the other hand, there is significant energy in relativistic
protons
(e.g. K $\sim$ 100, a value appropriate to the relativistic protons and
electrons observed at the top of the Earth's atmosphere)
and the X-rays are still SSC radiation, the energy density in relativistic
protons
must exceed that in the magnetic field. It is notable that the magnetic field
cannot be less than 1.5 $\times$ 10$^{-4}$ gauss since the SSC radiation would
then exceed the observed X-radiation.

The alternative is that B $>$ 1.5 $\times$ 10$^{-4}$ gauss in which case the 
predicted SSC emission would be too weak to account for the observed X-ray
emission. The X-rays would then have to be synchrotron radiation, as we
have recently argued for the jet and hot spot of Pictor A (Wilson, Young
\& Shopbell 2001). Synchrotron X-ray emitting electrons in hundreds of
microgauss strength magnetic fields have synchrotron half lives of order
years. They must thus be continuously accelerated within the hot spots.
Here, again, there are two options - direct electron acceleration to
$\gamma$ $\simeq$ 10$^{7}$ - 10$^{8}$ or a ``proton-induced cascade''.
In the latter process (e.g. Mannheim, Kr\"ulls \& Biermann 1991;
Biermann 1996), ultrahigh energy protons ($\gamma_{\rm p}$ $\sim$ 10$^{11}$),
perhaps shock accelerated, interact with photons to initiate a cascade, which
eventually yields relativistic electron-positron pairs. The peak of the 
radiation from this process is at $\sim$ MeV energies,
and the
spectral index predicted in the Chandra band
is $\alpha$ $\simeq$ 0.7
(Mannheim, Kr\"ulls \& Biermann 1991), in agreement with the spectral 
index observed. Observations
at higher energies than the Chandra band, where the
proton-induced cascade predicts a harder spectrum than the SSC model, are needed
to distinguish the two processes.

X-ray emission has now been detected from a number of hot spots in radio
galaxies. For three galaxies - 3C 123 (Hardcastle 2000), 3C 295 (Harris et al.
2000) and Cygnus A (present paper), the X-ray emission is well explained
by SSC emission from the radio synchrotron-emitting electrons with a magnetic 
field close to equipartition for K = 0. In three other galaxies - Pictor A
(Wilson, Young \& Shopbell 2001), 3C 120 (Harris et al. 1999) and 3C 390.3
(Harris, Leighly \& Leahy 1998) - the X-ray emission is orders of magnitude
too strong to be SSC emission with an equipartition magnetic field. Further,
the X-ray spectrum of Pictor A predicted by the SSC model disagrees with
that observed. In these cases, the X-rays are probably synchrotron
radiation from $\gamma$ $\sim$ 10$^{7}$ electrons. It is notable that 3C 295
and Cygnus A are in clusters with prominent cooling flows, while 3C 123 is in a
cluster with strong, extended X-ray emission and thus may be within a cooling
flow. In contrast, Pictor A, 3C 120 and 3C 390.3 are not in cooling flows.
It is tempting to speculate that the presence of high density surrounding gas 
may inhibit production of X-ray synchrotron emitting electrons
in hot spots. Several processes may be relevant to this effect, e.g. hot
spots propagating into the dense gas of a cooling flow should have lower
outward velocities, and there may be more entrainment of thermal gas into the
jet and hot spot. Such effects may reduce the efficiency with which
$\gamma$ $\sim$ 10$^{7}$ electrons are accelerated. Future
Chandra X-ray observations of additional radio galaxies should shed light on
this issue.

We are grateful to Rick Perley for providing the VLA 6 cm map of Cygnus A
in numerical form.
We also wish to thank the staff
of the Chandra Science Center, especially Dan Harris and Shanil
Virani, for their help.

\vfil\eject

\clearpage

\begin{deluxetable}{lllll}
\tablecolumns{5}
\tablewidth{0pc}
\tablecaption{Spectral fits to the X-ray emission of the
  hot spots\tablenotemark{a}}
\tablehead{
\colhead{Hot spot} & \colhead{$\rm N_{\rm H}$\tablenotemark{b}} &
\colhead{$\Gamma$} & \colhead{Normalization\tablenotemark{c}} &
\colhead{$\chi^2$ / d.o.f.}}
\startdata
A & $3.26^{+0.74}_{-0.61}$ & $1.84^{+0.22}_{-0.19}$ &
$\left(2.93^{+0.81}_{-0.63}\right)\times10^{-5}$ & 27 / 27 \\
D & $3.30^{+0.63}_{-0.26}$ & $1.75^{+0.19}_{-0.17}$ &
$\left(4.40^{+0.97}_{-0.77}\right)\times10^{-5}$ & 31 / 39 \\
\tablenotetext{a}{Errors are 90\% confidence for single parameter of
interest.}
\tablenotetext{b}{$10^{21}$ atom cm$^{-2}$.}
\tablenotetext{c}{Photons keV$^{-1}$ cm$^{-2}$
  s$^{-1}$ at 1\,keV.}
\enddata
\end{deluxetable}

\begin{deluxetable}{lllllll}
\tablecolumns{7}
\tablewidth{0pc}
\tablecaption{Parameters of synchrotron self-Compton models}
\tablehead{
\colhead{Hot spot} & \colhead{$B$} & \colhead{$n_{\rm e0}$} &
\colhead{$\alpha$} &
\colhead{$\gamma_{\rm break}$} & \colhead{$\gamma_{\rm max}$}&
\colhead{Radius} \\
\colhead{} & \colhead{(Gauss)} & \colhead{(cm$^{-3}$)} & \colhead{} &
\colhead{} & \colhead{} & \colhead{(kpc)}}
\startdata
A & $1.5\times10^{-4}$ & $1.9\times10^{-3}$ & 0.55 & 
$3.5\times10^3$ & $5.0\times10^4$ & 2.0 \\
D & $1.5\times10^{-4}$ & $5.5\times10^{-4}$ & 0.50 & 
$5.5\times10^3$ & $5.0\times10^4$ & 2.2 \\
\enddata
\end{deluxetable}

\vfil\eject

\clearpage

\figcaption[fig1.ps]
{A grey scale representation of the Chandra X-ray image of Cygnus A. The
shading is proportional to the square root of the intensity. Coordinates
are for epoch J2000.0 both here and in Fig. 2.
\label{Figure 1}}

\figcaption[fig2.ps]
{(a) X-ray emission (contours) superposed on a 6 cm VLA radio map (grey scale)
of the region of the western hot spots (A, the brighter, and B, $\simeq$
6$^{\prime\prime}$ SE of A). Contours are plotted at 2, 4, 8, 12, 16, 24 and 32
counts per pixel (0\farcs5 $\times$ 0\farcs5). The grey scale is proportional
to the square root of the radio brightness. (b)
As Fig. 2a, but for the eastern hot spots (D, the brighter, and E,
$\simeq$ 4$^{\prime\prime}$W of D). Contours are plotted at
2, 4, 8, 12, 16, 24, 32, 40, 48 and 56 counts per pixel.
\label{Figure 2}}

\figcaption[fig3.ps]
{(a) Spectrum of hot spot A. 
The points show the radio fluxes and the line through 
them the model of the synchrotron radiation. The ``bow tie'' is the Chandra
measured boundary of the X-ray spectrum (these error lines are 90\%
confidence after freezing N$_{\rm H}$ at its best fit value, which coincides
with the Galactic column). The solid line is the predicted SSC spectrum
for $\gamma_{\rm min}$ = 1 and the dashed line for
$\gamma_{\rm min}$ = 100. (b) As Fig. 3a, but for hot spot D.
The near infrared and optical upper limits
are from Meisenheimer, Yates \& R\"oser (1997), including their allowance
for obscuration.
\label{Figure 3}}

\clearpage

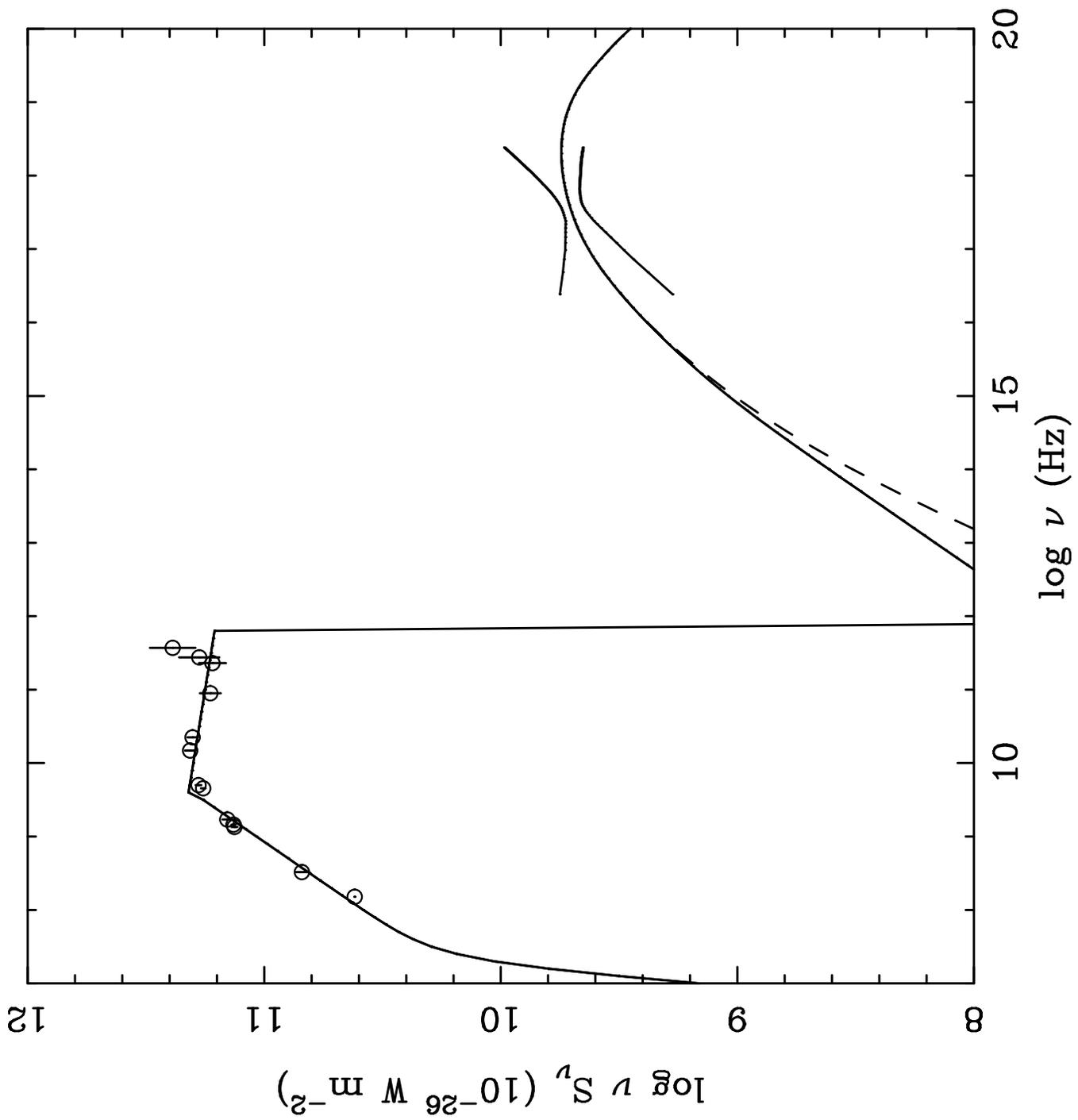
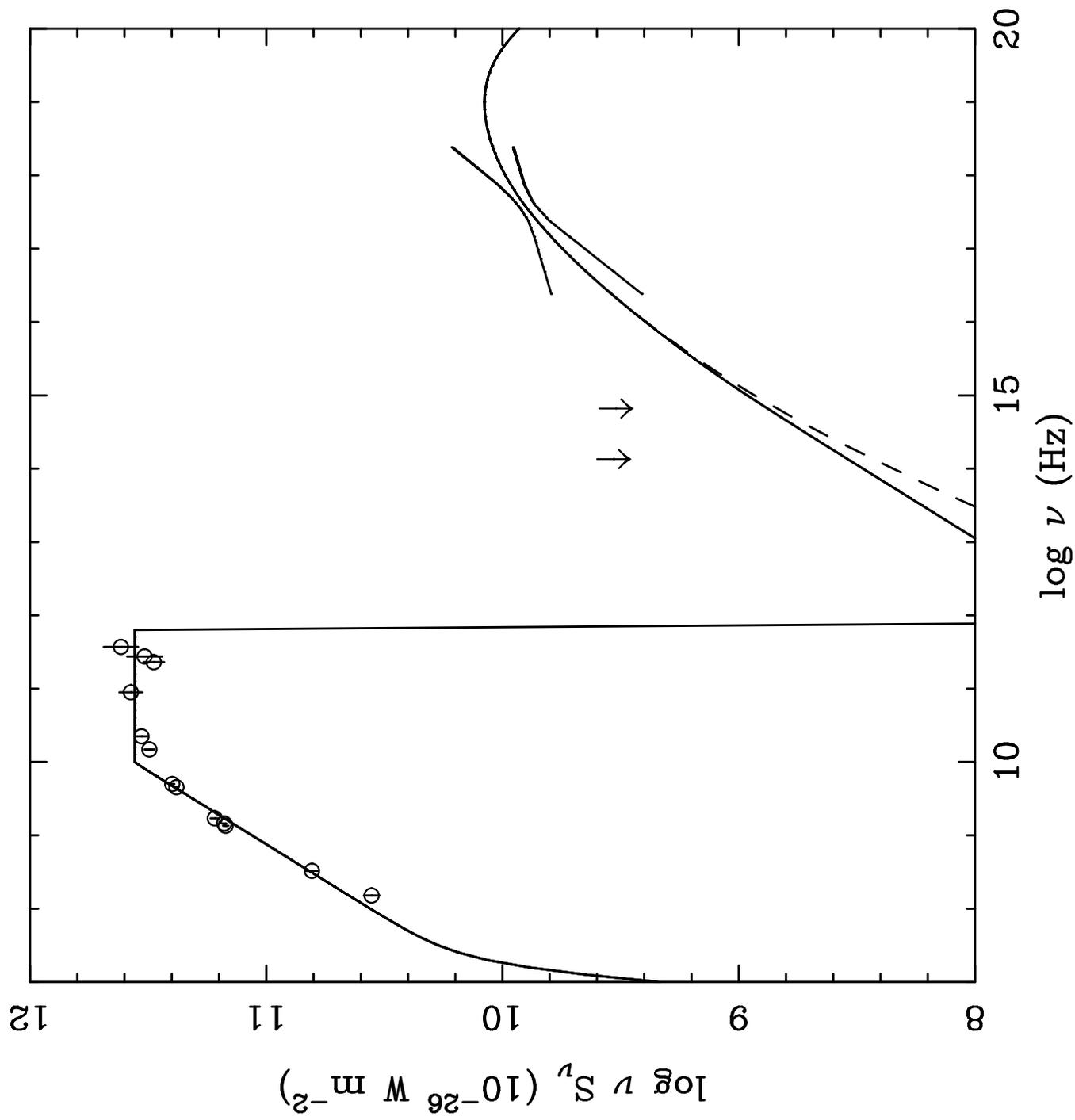
\end{document}